\documentclass[twocolumn,showpacs,superscriptaddress,amsmath,amssymb]{revtex4}

\usepackage{amsmath}
\usepackage{graphicx}
\usepackage{dcolumn}

\begin{document}

\title{Structure of particle-hole nuclei around $^{100}$Sn}
\author{L. Coraggio} 
\author{A. Covello} 
\author{A. Gargano} 
\author{N. Itaco}
\affiliation{Dipartimento di Scienze Fisiche, Universit\`a
di Napoli Federico II, \\ and Istituto Nazionale di Fisica Nucleare, \\
Complesso Universitario di Monte  S. Angelo, Via Cintia - I-80126 Napoli,
Italy}

\date{\today}

\begin{abstract}
We have performed shell-model calculations for the three odd-odd nuclei
$^{100}$In, $^{102}$In, and  $^{98}$Ag, with neutron particles and
proton holes around $^{100}$Sn. We have used a realistic effective
interaction derived from the CD-Bonn nucleon-nucleon potential, the
neutron-proton channel being explicitly treated in the particle-hole
formalism. Particular attention has been focused on the particle-hole
multiplets, which are a direct source of information on the neutron-proton
effective interaction.
We present our predictions for the two lowest lying multiplets in
$^{100}$In, for which no spectroscopic data are yet available.
For $^{98}$Ag and $^{102}$In comparison shows that  our
results are in very good agreement with the available
experimental data.
\end{abstract}


\maketitle

\section{Introduction}

One of the most interesting topics in nuclear structure is the study of nuclei near the limits 
of particle stability. Much attention is currently being focused on nuclei in the regions of shell closures 
off stability, in particular the $^{100}$Sn and $^{132}$Sn neighbors. These nuclei, unlike those 
in the $^{208}$Pb region, were until recently not accessible to spectroscopic studies. New data are now
becoming available for them, which  provide a challenging  testing ground for the effective interaction 
and  allow exploring  the evolution of the  
shell structure when approaching the proton and neutron drip lines.

To test the effective interaction between unlike nucleons, in a recent paper \cite{cor02a} we have 
studied nuclei with proton particles and neutron holes around $^{132}$Sn. 
This study has been performed
within the framework of the shell model making use of the CD-Bonn nucleon-nucleon 
($NN$) potential \cite{mac01}. 
We have considered $^{132}$Sn as a closed core
and derived the proton-neutron effective interaction explicitly in the
particle-hole ($ph$) formalism. 

Actually, a similar study \cite{kuo68} was performed for the heavier proton particle-neutron hole nucleus $^{208}$Bi more
than thirty years ago, employing $ph$ matrix elements derived from the Hamada-Johnston potential \cite{ham62}.
Since then, however, substantial progress has been made in both the development of high-quality 
$NN$ potentials and many-body  methods for deriving the effective interaction, which has stimulated our study 
of Ref. \cite{cor02a}. 

The results obtained in \cite{cor02a} are in good agreement with the experimental data, thus
evidencing the reliability of our proton particle-neutron hole matrix elements in the $^{132}$Sn region.
A relevant outcome of our study is that all multiplets in the two odd-odd nuclei
$^{132}$Sb and $^{130}$Sb show the same  peculiar behavior, which is consistent with the available experimental 
data. It is worth mentioning that the same behavior was  
evidenced for the multiplets in $^{208}$Bi \cite{moi69}, which was well reproduced by the 
calculation of Ref. \cite{kuo68}.

In this  paper, we present the results of a study of $^{100}$Sn neighbors performed  along the same lines of 
Ref. \cite{cor02a}. More precisely, we consider the three odd-odd nuclei
$^{100}$In, $^{98}$Ag, and $^{102}$In. Clearly, the most appropriate system to study the neutron 
particle-proton hole multiplets in this
mass region is $^{100}$In with only one neutron valence particle and one proton valence hole.
Although no experimental information is yet available for this nucleus, we have found it interesting 
to predict some of its spectroscopic properties to gain insight
into the effects of the neutron-proton effective interaction in the $^{100}$Sn region
also in the light of what we have learned from the study of nuclei around $^{132}$Sn. 
We hope that this may stimulate further experimental efforts to access this highly important nucleus.

Some experimental information is instead available for the two neighboring odd-odd nuclei 
$^{98}$Ag and $^{102}$In, which have two additional proton holes and  two additional neutrons 
with respect to $^{100}$In, respectively. The study of the $ph$ 
multiplets in these
nuclei not only is interesting in its own right, but
also provides a rather stringent test of the reliability of our predictions  for $^{100}$In. 

Preliminary results for $^{98}$Ag and $^{102}$In have been already presented in Refs. \cite{cov03,cov04}. 
In Ref. \cite{cov04} we have also reported some results for $^{208}$Bi, which turn out to be in better agreement
with experiment than those of the early realistic calculations of Ref. \cite{kuo68}.

The outline of the paper is as follow. In Sec. II we give a brief description of our calculations
focusing attention  on the choice of the neutron single-particle (SP) and the proton single-hole (SH)
energies. Our results are presented and compared with the experimental data in Sec. III.
Section IV contains a discussion and a summary of our conclusions. 

\section{Outline  of calculations}

We assume that $^{100}$Sn is a closed core and let the valence
neutrons occupy the five levels $0g_{7/2}$, $1d_{5/2}$, $1d_{3/2}$, $2s_{1/2}$,
and $0h_{11/2}$ of the 50-82 shell, while for the proton
holes the model space includes the four levels $0g_{9/2}$, $1p_{1/2}$,
$1p_{3/2}$, and $0f_{5/2}$ of the 28-50 shell. As input to our
shell-model calculations we need the neutron SP and the proton SH
energies as well as the two-body matrix elements of the effective interaction.
As regards the latter, the calculation for
$^{100}$In requires only the neutron-proton matrix elements in the
$ph$ formalism, while for $^{98}$Ag and $^{102}$In
we also need proton-proton and neutron-neutron matrix elements in the
hole-hole ($hh$) and particle-particle ($pp$) formalism, respectively.

As already mentioned in the Introduction, our 
effective interaction has been derived from the CD-Bonn $NN$ potential \cite{mac01}.
The difficulty posed by the strong short-range repulsion contained in the bare
potential has been overcome by constructing a renormalized
low-momentum potential, $V_{\rm low-k}$, that preserves the physics
of the original potential up to a certain cut-off momentum $\Lambda$.
The latter is a smooth potential that can be used
directly in the calculation of shell-model effective interactions.
A detailed description of our derivation of $V_{\rm low-k}$ can be found
in Ref. \cite{bog02}. In the present paper, we have used  for  $\Lambda$ the value 2.1 fm$^{-1}$.

Once the $V_{\rm low-k}$ is obtained, the calculation of the effective
interaction is carried out within the framework of  the $\hat {Q}$-box plus folded diagram method
\cite{kre80}. In the calculation of $\hat{Q}$ we have included diagrams up
to second order in $V_{\rm low-k}$. A description of the derivation of the
effective interaction 
in the $pp$ and $hh$ formalism can be found in Refs. \cite{jia92} and \cite{cor00},  respectively, 
while the matrix elements in the $ph$ formalism have been
explicitly derived in Ref. \cite{cor02a}.

As regards the neutron SP and proton SH energies,
they cannot be taken from experiment, since no
spectroscopic data are yet available  for $^{101}$Sn and $^{99}$In.
In two previous papers, we have determined them by an 
analysis of the low-energy spectra of the odd Sn isotopes with $A \leq
111$ for the former \cite{and96} and of the $N=50$ isotones with $ A \geq 89$ 
for the latter \cite{cor00}.
Recently, however, the $\frac{5}{2}^+$ ground state and the first-excited $\frac{7}{2}^+$ state 
have been  identified in $^{103}$Sn \cite{fah01}. 
We have therefore found it appropriate  to include these states in our analysis of 
the Sn isotopes, which yields the value of 0.010 MeV for 
$\epsilon_{d_{5/2}} - \epsilon_{g_{7/2}}$, as compared to
the value of -0.200 MeV adopted in Ref. \cite{and96}.
For the sake completeness, the values of the neutron SP
and proton SH  energies adopted in the present
calculation are reported in Table I.

\begin{table}[ht]
\caption{Proton single-hole and neutron single-particle energies (in MeV).}
\begin{ruledtabular}
\begin{tabular}{c|c|c|c}
$\pi(n,l,j)^{-1}$&$\epsilon$ & $\nu (n,l,j)$& $\epsilon$\\
\colrule
$0g_{9/2}$ & 0    & $0g_{7/2}$ & 0 \\
$1p_{1/2}$ & 0.700   & $1d_{5/2}$ & 0.010 \\
$1p_{3/2}$ & 2.100   & $2s_{1/2}$ & 2.200 \\
$0f_{5/2}$ & 3.100  & $1d_{3/2}$ & 2.300  \\
&   &                $0h_{11/2}$ & 2.700  \\
\end{tabular}
\end{ruledtabular}
\end{table}

To conclude this section, it is worthwhile to comment on 
the $\nu d_{5/2} - \nu g_{7/2}$ spacing. From the experiment we know that the $N=51$
isotones with $38 \leq Z \leq 46$ are characterized by a $\frac{5}{2}^{+}$ ground state
with the excitation energy of the $\frac{7}{2}^{+}$ state going down when approaching 
the proton shell closure. Furthermore, the  
data for the light Sn isotopes indicate that the  $d_{5/2}$ and $g_{7/2}$ 
levels should lie close in energy,
no subshell effect being  observed as evidenced by the
constancy of the excitation energy of the first $2^+$ state up to $^{110}$Sn.
However, the spacing between these two levels cannot be 
firmly established from the available experimental data and it is still an open question the ordering of the $g_{7/2}$   
and $d_{5/2}$ level in $^{101}$Sn.
   
The SP energies of the $\nu d_{5/2}$ and $ \nu g_{7/2}$ levels adopted in this study
are almost degenerate, with the $ \nu g_{7/2}$ level lying only 10 keV
below the $\nu d_{5/2}$ one. It should be pointed out, however,  that this choice 
is based on the above mentioned analysis of the light Sn isotopes with $A \geq 103$. It 
therefore depends on the $J^{\pi}=0^{+}$ neutron-neutron matrix elements 
of the effective interaction. On these grounds, we consider that our value may be subject to
an uncertainty of the order of 100 keV.

\section{Results}

We now present the results of our calculations for the neutron-proton multiplets in $^{100}$In, $^{98}$Ag, and
$^{102}$In and compare them with the experimental data. All calculations have 
been performed using the OXBASH shell-model code \cite{oxb}. 

\begin{figure}[ht]
\includegraphics[scale=1.0,angle=0]{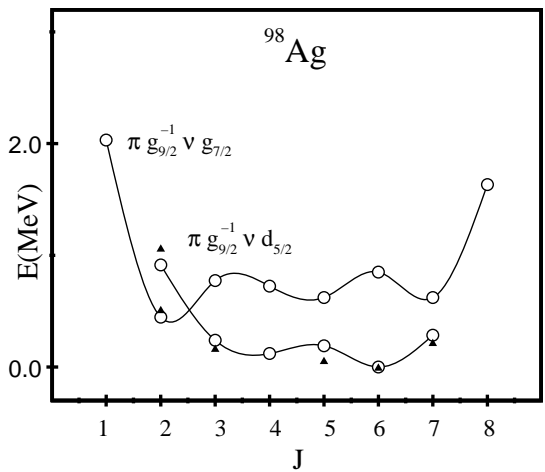}
\caption{Neutron particle-proton hole multiplets in $^{98}$Ag. The
  calculated results are represented by open circles while the
 experimental data by solid triangles. The
lines are drawn to connect the points.} 
\end{figure}

Let us start with $^{100}$In. First we note that the wave functions for low-lying states 
have little configuration mixing and can be then grouped into multiplets.  
In Fig. 1 we show our predictions for the $\pi g^{-1}_{9/2} \nu d_{5/2}$ and  $\pi g^{-1}_{9/2} \nu g_{7/2}$ 
multiplets. For both of them 
the weight of the leading component in each member is at least 77\%, the remaining percentage  in the states 
of the $\pi g^{-1}_{9/2} \nu d_{5/2}$ multiplet coming
from the $\pi g^{-1}_{9/2} \nu g_{7/2}$ configuration and vice versa.

From Fig. 1 we see that the two multiplets have a similar shape with a splitting between the centroid energies of about
400 keV, the centroid of the $\pi g^{-1}_{9/2} \nu g_{7/2}$ multiplet being the highest one. Since
the $g_{7/2}$ and $d_{5/2}$ neutron levels are almost degenerate, this splitting can 
be only attributed to the difference between the average energy contribution of the neutron-proton interaction
for the two different configurations. This implies therefore that the matrix elements of the effective interaction, 
which is essentially repulsive 
in the particle-hole channel, are generally  stronger for the states of the  $\pi g^{-1}_{9/2} \nu g_{7/2}$
multiplet.
This is also characterized by  a larger dispersion ($\Delta= 630$ keV)
as compared to that ($\Delta=360$ keV) of the $\pi g^{-1}_{9/2} \nu d_{5/2}$ multiplet.

As regards the shape of the two multiplets, we see from Fig. 1 that it is a main feature  of both of them 
that
the states with the minimum and maximum $J$ have the highest excitation energy, while the state with next to the
highest $J$ is the lowest one.  This reflects the very small contribution of the effective interaction
(either  attractive or repulsive)
to the $J_{\rm max}-1$ state, the largest repulsive effect occurring for 
the $J_{\rm max}$ and $J_{\rm min}$ states. It is just this feature that causes 
the first and last states of the
$\pi g^{-1}_{9/2} \nu d_{5/2}$ multiplet to lie above the 
$2^+$ and $7^+$ members of the   $\pi g^{-1}_{9/2} \nu g_{7/2}$  multiplet. As can be seen from Fig. 1,
all the other members of the $\pi g^{-1}_{9/2} \nu d_{5/2}$ multiplet are instead yrast states. 
It is worth mentioning that the configuration mixing  also contributes to further split 
the two $2^+$ and the two $7^+$ states .

The peculiar behavior exhibited  by the proton hole-neutron multiplets in $^{100}$In turns out to be 
quite similar to that of the  multiplets in the proton 
particle-neutron hole nuclei $^{132}$Sb  and $^{208}$Bi, which we have discussed in the Introduction.  
This pattern appears to be typical of the particle-hole multiplets 
in odd-odd nuclei irrespective of the mass region and the nature of the hole, either
a proton or a neutron. 
This similarity    
makes evident  that the main features of the effective interaction in the 
$ph$ channel are essentially the same in the $^{100}$Sn, $^{132}$Sn, and $^{208}$Pb regions. 
Note that we have found that no significant role is played by the configuration mixing
and therefore the behavior of each multiplet is essentially determined by the matrix elements of the
effective interaction for the corresponding configuration.

\begin{figure}[ht]
\includegraphics[scale=1.0,angle=0]{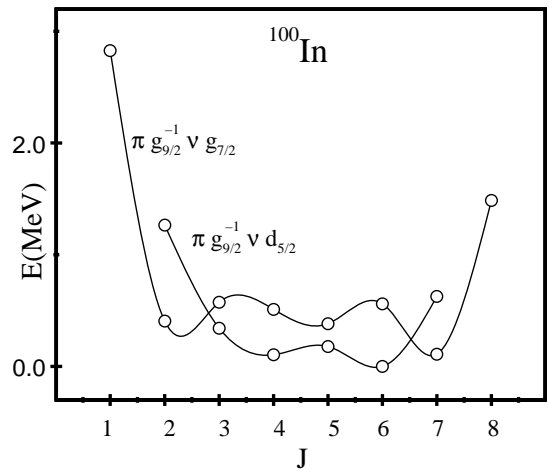}
\caption{Calculated neutron particle-proton hole multiplets in $^{100}$In. The
lines are drawn to connect the points.}
\end{figure}

We now would like to draw attention to our results as regards the spin and parity
of the ground state.  The nature of this state is of great interest since it may provide
information  on the $g_{7/2} - d_{5/2}$ neutron spacing \cite{gra02}.
We predict the ground state of  $^{100}$In  to be the 
$J^\pi = 6^+$ member of  the $\pi g^{-1}_{9/2} \nu d_{5/2}$
multiplet, while  the $J^\pi = 7^+$  state, which originates  from the $\pi g^{-1}_{9/2} \nu g_{7/2}$ configuration,
lies at 110 keV excitation energy. 
We have found that to obtain the  latter as ground state the $\nu g_{7/2}$ level should go at least 200 keV
below the $ \nu d_{5/2}$ one. This value is needed to compensate for the
effect of the repulsive effective interaction, which, as mentioned above, 
is stronger in the $\pi g^{-1}_{9/2} \nu g_{7/2}$ configuration. However,
based on our previous discussion (see Sec. II), such a spacing 
seems to be too large.

Let us now come to $^{98}$Ag and $^{102}$In 
which are the odd-odd nuclei closest to $^{100}$Sn for which some experimental information is available. The calculated
multiplets $\pi g^{-1}_{9/2} \nu g_{7/2}$ and $\pi g^{-1}_{9/2} \nu d_{5/2}$ for  $^{98}$Ag and $^{102}$In are reported
in Figs. 2 and 3, respectively, where they are compared with the experimental data \cite{nndc}.  
We have identified as members of the 
multiplets the states dominated by the corresponding $ph$ configuration with the two
remaining proton holes or neutron particles forming a zero-coupled pair.   
In  both nuclei we have found that the calculated yrast and yrare states with angular momentum from 2 to 7 belong
to one of the two
multiplets, while the $8^+$ member of the  $\pi g^{-1}_{9/2} \nu g_{7/2}$ multiplet turns out to be the
second excited $8^+$ state.
As for the $1^+$ member of the latter multiplet, we find that it is 
the third  $1^{+}$ state in $^{98}$Ag and the fourth  one   
in $^{102}$In.

\begin{figure}[ht]
\includegraphics[scale=1.0,angle=0]{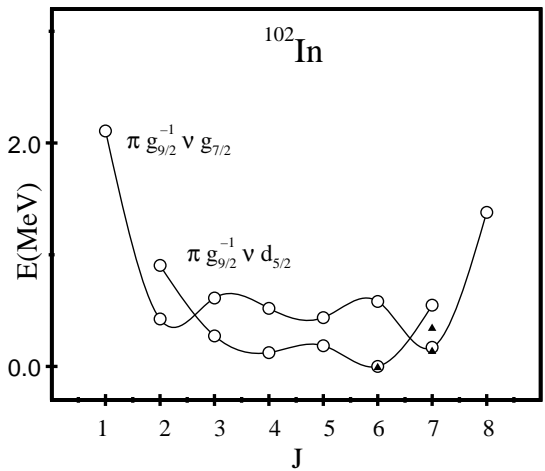}
\caption{Same as Fig. 2, but for $^{102}$In.}
\end{figure}

It is worth mentioning that the percentage of components other than those characterizing  the
multiplets is rather large in all states reported in Figs. 2 and 3, exceeding 50\% for some of them. 
We have verified that this mixing is rather sensitive to the spacing between the
$d_{5/2}$ and $g_{7/2}$ neutron levels. For instance, if one takes the  $\nu g_{7/2}$  level
200 keV above the  $\nu d_{5/2}$ 
one, the two $2^+$, as well as the two $7^+$, states in $^{102}$In change their nature, 
namely  the yrast state become a member the $\pi g^{-1}_{9/2} \nu d_{5/2}$ multiplet while the yrare a member of   
the  $\pi g^{-1}_{9/2} \nu g_{7/2}$ multiplet. However, the pattern of each  multiplet in both nuclei
is on the whole not affected by changes in the
$g_{7/2} - d_{5/2}$ neutron spacing, remaining quite similar to that of the corresponding 
multiplet in $^{100}$In.
 
Let us now come to the comparison with the experimental data. In this connection, some comments on the values reported in 
Figs. 2 and 3 are in order. 
As regards $^{98}$Ag, we have included the first six low lying states, assuming the 
$(2^{+},3^{+})$ and $(2,3^{+})$ levels at 0.515 and 1.066 MeV, respectively,  to have both $J^{\pi}=2^+$ as
suggested by our calculations.  
We have not included the experimental $8^+$ state at 1.115 MeV, which has been identified with our
first calculated $8^+$ state at 1.315 MeV. It should also be noted  that none of the experimental $1^+$ states has been 
reported, since no  safe identification can be made with our calculated $1^+$ state reported in Fig. 2.  
The experimental data shown in Fig. 3 for $^{102}$In include only three states: the ground state for which we have adopted 
 $J^{\pi}=6^+$  according to Refs. \cite{sew95,lip02,soh02}, and the two first excited states both with $J^{\pi}=7^+$. The  $8^+$
at 0.980 MeV has not been reported since, as for $^{98}$Ag, it has been identified with our first $8^+$
at 1.210 MeV. 
From Figs. 2 and 3 we see that the calculated energies are in good agreement with the experimental ones, the largest 
discrepancy being 200 keV for the second $7^+$ state in  $^{102}$In.

We conclude by noting that our calculations predict for the ground state of the three nuclei 
considered $J^{\pi}= 6^+$. We have verified that 
this result is independent of reasonable changes in the $g_{7/2} - d_{5/2}$ neutron spacing.

\section{Summary and conclusions}
In this work, we have performed shell-model calculations for
$^{100}$In, $^{98}$Ag, and $^{102}$In, which are
the immediate odd-odd neighbors of doubly magic $^{100}$Sn.
The effective interaction has been derived from the CD-Bonn
nucleon-nucleon potential \cite{mac01}, making use of a new approach \cite{bog02}
to the renormalization of the short-range repulsion of the nucleon-nucleon potential.
The main aim of this work has been to study the particle-hole multiplets in
this region to obtain information on  the neutron-proton effective interaction.

We have found that the proton hole-neutron multiplets in the
three above nuclei exhibit the same behavior as that of the multiplets in the
proton particle-neutron hole nuclei around doubly magic $^{208}$Pb and
$^{132}$Sn.
Namely, the highest- and lowest-spin members of each
multiplet have the highest excitation energy, while the state with
next to $J_{\rm max}$ is the lowest, the latter feature being in agreement 
with the predictions of the Brennan-Bernstein coupling rule \cite{bre60}.
This pattern,  which is  essentially independent of
the mass region, can be  directly related to the effective interaction between unlike nucleons
in the particle-hole channel. In fact, the diagonal matrix elements of the effective interaction for a given 
configuration show the same behavior, the configuration mixing, 
as mentioned in Sec. III, playing a minor role. In this connection, we would like  to
mention the work of Ref. \cite{soh02}, where empirical neutron particle-proton hole matrix 
elements have been used. These  
have been determined by fitting the energies of 150 levels in 10
nuclei with $46 \leq Z \leq 49$  and   $51 \leq N \leq 54$. In Ref. \cite{soh02} the diagonal matrix elements
for the $\pi g_{9/2}^{-1} \nu g_{7/2}$, $\pi g_{9/2}^{-1} \nu d_{5/2}$, and
$\pi g_{9/2}^{-1} \nu h_{11/2}$ configurations are reported and  it is interesting to note that
their behavior is somewhat different from that of our realistic matrix elements. In fact, while 
the $J_{\rm min}$ and $J_{\rm max}$ matrix elements still  have the largest values, the matrix
element with $J= J_{\rm max}-1$ is not the smallest one, except  for the 
$\pi g_{9/2}^{-1} \nu d_{5/2}$ configuration. 

To conclude this discussion, it is worth mentioning  that  
in Refs. \cite{cov04,gar04} we have reported the diagonal matrix elements 
of our effective interaction for some configurations in
$^{100}$In, $^{208}$Bi, and $^{132}$Sb.  
From an  analysis of these matrix elements it turned out that 
the renormalizations of the $V_{\rm low-k}$ potential due to core-polarization processes, although not very large in magnitude, 
are quite relevant for the pattern of the particle-hole multiplets.
 
As regards the comparison with experiment,  we have shown that the data available for the two odd-odd nuclei 
$^{98}$Ag
and $^{102}$In are well reproduced by our calculations providing confidence in our 
predictions  for $^{100}$In. Finally, we would like to point out that for the ground state
of this nucleus we predict $J^{\pi} = 6^+$ originating from
the $\pi g_{9/2}^{-1} \nu d_{5/2}$ multiplet. This result turns out to be
quite insensitive to reasonable changes in the $g_{7/2} -d_{5/2}$ neutron spacing.

\begin{acknowledgments}
This work was supported in part by the Italian Ministero
dell'Istruzione, dell'Universit\`a e della Ricerca  (MIUR).
\end{acknowledgments}

\end{document}